\documentclass{aastex}

\usepackage{emulateapj5} 

\newcommand{\Ga}{\Gamma}

\newcommand{\ten}[2]{$#1\times 10^{#2}$}
\newcommand{\asca}{{\sl {ASCA\/}}}

\newcommand{\euve}{{\sl {EUVE\/}}}
\newcommand{\sax}{{\it BeppoSA$\!$X}}

\newcommand{\tcycle}{t_{\rm cycle}}
\newcommand{\tcol}{t_{\rm col}}
\newcommand{\tshot}{t_{\rm shot}}
\newcommand{\tchr}{t_{\rm chr}}
\newcommand{\tcrs}{t_{\rm crs}}
\newcommand{\tang}{t_{\rm ang}}

\newcommand{\vs}{v_{\rm s}}

\newcommand{\gm}{\Gamma _{\rm m}}
\newcommand{\gfs}{\Gamma _{\rm fs}}
\newcommand{\grs}{\Gamma _{\rm rs}}

\newcommand{\betars}{\beta_{\rm rs}}

\newcommand{\Em}{E_{\rm m}}
\newcommand{\gav}{\Gamma_{\rm avrg}}
\newcommand{\dg}{\Delta \Gamma}
\newcommand{\sg}{\sigma_{\Gamma}}
\newcommand{\sgd}{\sigma_{\Gamma} '}
\newcommand{\sld}{\sigma_{l} '}

\newcommand{\st}{\sigma_{\rm T}}
\newcommand{\gmax}{\gamma_{\rm max}}
\newcommand{\nmax}{\nu_{\rm max}}
\newcommand{\rfo}{r_{\rm fo}}

\newcommand{\Ub}{U_{\rm B}}
\newcommand{\usoft}{U_{\rm soft}}
\newcommand{\me}{m_{\rm e}}
\newcommand{\bshock}{\beta_{\rm s}}

\slugcomment{to appear in ApJ}

\shorttitle{Tanihata et al.}
\shortauthors{Implications of Variability Patterns of TeV Blazars}

\begin{document}


\title{Implications of Variability Patterns observed in TeV Blazars on the Structure of the Inner Jet
}


\author{Chiharu Tanihata\altaffilmark{1,2}, 
Tadayuki Takahashi\altaffilmark{1,2},
Jun Kataoka\altaffilmark{3}, and
Greg M. Madejski\altaffilmark{4}
}


\altaffiltext{1}{Institute of Space and Astronautical Science,
        3-1-1 Yoshinodai, Sagamihara, 229-8510, Japan}
\altaffiltext{2}{Department of Physics, University of Tokyo,
        7-3-1 Hongo, Bunkyo-ku, Tokyo, 113-0033, Japan}
\altaffiltext{3}{Department of Physics, Tokyo Institute of Technology,
        Tokyo, 152-8551, Japan}
\altaffiltext{4}{Stanford Linear Accelerator Center, 
        Stanford, CA, 94309-4349, USA}

\begin{abstract}
The recent long look X--ray observations of TeV blazars have
revealed many important new features concerning their 
time variability. 
In this paper, we suggest a physical interpretation
for those features based on the framework of the internal and
external shock scenarios.  
We present a simplified model applicable to TeV blazars, 
and investigate through simulations how each of the model parameters 
would affect to the observed light curve or spectrum.
In particular, we show that the internal shock scenario 
naturally leads to all the observed variability properties
including the structure function, but for it to be applicable, 
the fractional fluctuation of the initial
bulk Lorentz factors must be small, $\sgd \equiv \sg / \gav \ll 0.01$.
This implies very low dynamical efficiency of the internal shock scenario.
We also suggest that several observational quantities -- such as the 
characteristic time scale, the relative amplitude of flares 
as compared to the steady (``offset'') component, and the 
slope of the structure function -- can be used to probe 
the inner jet.  The results are applied to the TeV blazar Mrk~421, 
and this, within the context of the model, leads to the determination 
of several physical parameters:  
the ejection of a shell with average thickness of 
$\sim10^{13}$ cm occurs on average every 10 minutes, 
and the shells collide $\sim10^{17}$ cm away from the central source.
\end{abstract}




\section{INTRODUCTION}
Blazars are active galactic nuclei exhibiting the most rapid 
and largest amplitude variability of all AGN. 
Historically, radio observations first revealed that
the emission was luminous and rapidly variable.  With 
that, assuming that the radio emission was due to synchrotron 
radiation, calculations of Compton up-scattering predicted much 
higher X--ray fluxes than the observed values unless the radio 
emission was relativistically beamed \citep{hoyle66,jones74}.
This led to our current model for blazars where the entire 
electromagnetic emission arises in a relativistic jet pointing 
close to the line of sight \citep{blandford78}. 
Subsequent radio observations using the Very Long Baseline 
Interferometry (VLBI) showed superluminal motion in many sources,
which served as the direct evidence for the relativistic motion.

The broadband spectra of blazars consist of two peaks, 
one in the radio to optical--UV range (and in some cases, reaching to
the X--ray band), and the other in the X--ray to $\gamma$--ray region.
From the high polarization of the radio to optical emission,
the lower energy peak is best interpreted as produced 
via the synchrotron process by relativistic electrons in the jet.
The higher energy peak is believed to be due to Compton up-scattering 
by the same population of relativistic electrons.  Several
possibilities exist for the source of the seed photons;  these can 
be the synchrotron photons internal to the jet 
\citep{jones74,ghisellini89}, 
but also external, such as from the broad emission line clouds
\citep{sikora94} or from the accretion disk \citep{dermer92,dermer93}.

Blazars are commonly detected as $\gamma$--ray sources.  A number of 
them with peak synchrotron output in the X--ray range also have been 
detected in the TeV range with ground-based Cherenkov arrays. 
These are the so-called ``TeV blazars.''  In TeV blazars, X--rays 
provide the best means for studying variability properties:  this 
is because X-ray flux is presumably produced by electrons that are 
accelerated to the highest energies (where the cooling time scales 
are most rapid), and thus the dilution by the non-varying components 
is the smallest.  

Variability studies of blazars have entered a new stage
after a number of continuous long-look X--ray observations 
conducted with the \asca\ satellite.  One such observation, 
of Mrk~421, conducted in 1998, showed for the first time that 
flaring is actually occurring on a daily basis, and that 
long-duration flares detected in previous observations
were probably unresolved superpositions of multiple, more rapid 
flares \citep{tad00}.  Such excellent data 
provided the new knowledge about the radiation processes,
allowing an exploration of the actual dynamics of the
particles that are accelerated in the jets.

In \S2, we present the properties of X--ray variability observed 
from TeV blazars, and summarize the issues we need to explain.
In \S3, we consider the internal shock scenario,
which involves shells propagating rapidly along the jet.  
Using this scenario, we simulate expected X-ray light curves,   
and study how various quantities that 
can be measured from observations depend on the input 
parameters to the model.  In \S4, we compare the 
results of simulations to the data to determine 
whether the observed variability features can be
reproduced, and if so, what are the implications on model parameters. 
We briefly consider the possibility of the external shock 
scenario in \S5, and give a summary in \S6.

\section{VARIABILITY PROPERTIES OBSERVED IN TEV BLAZARS}
\label{sec:obs}
One of the most surprising results from the long look
observations of TeV blazars was the 
repeated occurrence of flares with a time scale of $\sim$ one day.
This was first observed in the 7--day observation
of Mrk~421 in 1998 \citep{tad00}, and
was confirmed via the 10--day observations of
both Mrk~501 and PKS~2155--304 in 2000 \citep{tanihata01}.
The latter two sources were in a relatively low flux state
compared to previous observations, which implied that 
a high state is not a requirement for rapid variability.
In other words, this feature indicates that
the time scale of the rise and decay of the flares
are similar to the time scale of the repeating of the flares.

Another observational fact is that the X--ray flares always 
appear to lie on top of an underlying offset-like component 
(see, e.g., light curves in Urry et al. 1997; Tanihata et al. 2001;
Zhang et al. 2002).
At this point, we cannot distinguish whether this
is due to superpositions (or pile-ups) of flares,
or whether there is a separate, non-varying offset component,
but the observations indicate that there is always some
component such that a flare does not start from zero,
but from a level comparable to the flare amplitude.

An important advantage of long continuous observations is that
the variability can be treated statistically.
Recent structure function analysis have showed that
there is clearly a characteristic time scale $\tchr$
of an order of a day for each of three TeV blazars, 
Mrk~421, Mrk~501, and PKS~2155--304,
and also that the variability power at shorter
time scales than $\tchr$  is strongly suppressed
\citep{kataoka01,tanihata01}.  
Together with the near-symmetry of the flares and
also by comparing the various possible time scales, 
it was suggested that this $\tchr$
is determined by some dynamical time scale,
instead of energy dependent cooling or acceleration times
\citep{kataokaD}.

X--ray emission from TeV blazars shows spectral variability, where 
the common trend indicates a harder X--ray spectrum for higher 
intensity states.  Furthermore, the improved spectral coverage allows 
a measurement of the exact location of the synchrotron peak.  Perhaps 
the most striking case is Mrk~501, where the peak frequency shifted 
from below 1 keV to over 100 keV during the very high flux state 
\citep{pian98,fab01}.  Likewise, \citet{fossati00b} has shown using the 
\sax\ observations of Mrk~421 in 1997 and 1998 that the 
peak frequency shifted to the higher energy for higher intensity states;  
this is particularly apparent during the flare observed in 1998 April.
This was also shown in the long look observation of Mrk~421
in 1998 using the combined \asca\ and \euve\ spectrum \citep{tanihata02}.
Through a detailed analysis of the spectral evolution at the 
rise of the flare, we found several flares that start to appear from the 
higher energy, which strongly suggests that an appearance of 
a new harder (i.e., higher synchrotron peak frequency)
emission component is associated with the generation of the flare.

Summarizing, the following features are the 
issues concerning the variability derived from observations,
that we need to explain in considering any model:  
(1) Daily-flares (i.e. $\tcycle \sim \tchr$),
(2) the offset component, 
(3) the structure function, and
(4) the energy dependence.

\section{THE INTERNAL SHOCK SCENARIO}
\label{sec:sim}
Among many studies addressing the mechanism of particle acceleration 
in jets, acceleration via shocks appears to be the most viable.
Such shocks can efficiently accelerate particles to very high energies
(e.g. Longair 1994; Bell 1978; Drury 1983;
Blandford \& Eichler 1987; Jones \& Ellison 1991),
and since it is highly probable that shocks form inside jets, we
consider it in more detail here.  

In order to form a shock, there must be a large velocity difference 
between the colliding parcels of matter.  
The key idea of the internal shock model is a shock-in-jet scenario, 
where the central engine injects energy into the jet in a
discontinuous manner, producing individual shells having
slightly different bulk Lorentz factors and energies.
If this occurs, there will be collisions 
by a faster shell catching up to a slower one, forming a shock.
This internal shock scenario is among the most promising models 
to explain the emission of gamma--ray bursts (e.g. Sari \& Piran 1995),
although it was originally suggested in reference to AGN 
more than 20 years ago by \citet{rees78}.  It has been recently 
suggested that this model could be successfully applied to blazars, as 
it can explain some of their basic properties such as the low 
efficiency \citep{gab01,spada01,sikora01}.

\subsection{The Model}
In order to investigate whether the observed variability 
properties can be reproduced in the internal shock scenario,
we developed a simulation code. Recently, \citet{spada01} has 
given a detailed calculation covering the formation, 
propagation, and collision of such shells.  Included are hydrodynamic
calculations to determine the structure of the shock fluid, 
and the full radiation spectrum is derived by summing up 
all the locally produced spectra from the electrons accelerated
by the shocks.  The model and simulations presented here are a 
much simplified version, to be compared specifically against the 
actual measurements of TeV blazars.  

Due to the difference of the initial velocity of the 
shells ejected from the base of the jet, collisions 
occur when a faster shell catches up to a slower one. 
This is where the shock is formed, electrons are accelerated,
and lose energy through radiation.  Since we consider only 
the colliding shells, in our simulations we generate pairs 
of shells with one having a bulk Lorentz factor (BLF) of $\Ga_1$ 
ejected from the base of the jet, and a following shell 
with a BLF of $\Ga_2$ ejected after an initial separation 
distance of $D_0$.  If $\Ga_2$ is larger than $\Ga_1$, the latter
shell will catch up to the former one, and the two will merge 
into a single shell producing a shock.  Each collision then 
generates radiation (called hereafter a ``shot'') for a duration 
of $\tshot$.  The time profile of each shot is assumed to have a 
symmetric linear rise and decay.  The collisions are distributed 
randomly in time following a Poisson distribution, and superposition 
of these individual shots results in the output light curve.

For simplicity, the rest mass of all shells is assumed to be the same, 
and the shell thickness $l$ is assumed to be equal to the
initial separation $D_0$.  The average frequency of the collision 
is set to be consistent with the separation of the shells
(i.e. $F_{\rm col} = c/2(D_0+l)$), and only the first collision 
is considered.  The initial BLFs are assumed to be distributed around 
an average value $\Ga_{\rm avrg}$, following a Gaussian distribution 
with its width described by the sigma, $\sigma_\Ga$.  In this case, 
there are only three input parameters:  two of them describe the 
distribution of the initial BLFs of the ejected shells ($\gav$, $\sg$), 
and one describes the initial separation of the two colliding shells 
($D_0$).  We also define the fractional width $\sgd$ $\equiv \sg / \gav$.

With the assumptions described above, using the momentum and energy 
conservation laws in an inelastic collision, the BLF of the merged shell
is 
\begin{equation}
\gm =(\Ga_1\Ga_2)^{1 \over 2}.
\label{eq:d_is:gm}
\end{equation}
The newly generated total internal energy is 
\begin{equation}
E_{\rm m}=Mc^2(\Gamma _1-\Gamma _{\rm m})+Mc^2(\Gamma _2-\Gamma _m),
\label{eq:d_is:em}
\end{equation}
and thus the dynamical efficiency is given by
\begin{equation}
\eta=\frac{Mc^2(\Gamma _1-\Gamma _{\rm m})+Mc^2(\Gamma _2-\Gamma _m)}{Mc^2
\Gamma_1 + Mc^2 \Gamma_2}.
\label{eq:d_is:eta}
\end{equation}
We assume that all of the newly generated internal energy is converted 
to the random energy of the electrons.

Each collision will take place at distance
\begin{equation}
D\sim \frac{2\Gamma_1 ^2 \Gamma_2 ^2}{\Gamma_2^2-\Gamma_1^2} D_0 
\label{eq:d_is:d}
\end{equation}
from the core, where in making the above approximation, we 
used $\Gamma_1^2$, $\Gamma_2^2 \gg 1$. 
We assume that the jet is
collimated into a cone with an opening angle
$\theta \sim 1/\Ga$. The radius of the shell
at the location of the collision can thus be written as
\begin{equation}
R=D\tan \theta \sim {D \over \Ga}.
\label{eq:d_is:r}
\end{equation}

The time duration of each shot $\tshot$ results from
several competing time scales:  the acceleration and cooling 
time scales, and the dynamic time scale, which 
includes the hydrodynamic time and the angular spreading time.
It has been shown for TeV blazars that the cooling or 
acceleration times are 
significantly shorter than the observed rise and 
decay time scales of the 
X--ray and gamma--ray flares \citep{kataokaD,tanihata01},
and thus the dynamical time scale most likely 
determines $\tshot$, and also filters out any faster variability.

The hydrodynamic time scale is determined by the
time that the shock takes to cross the shell.
Using the conservation of mass, energy, and momentum
at the shock, and the equality of pressure and 
velocity along the contact discontinuity, the
Lorentz factors of the forward and reverse shocks
$\gfs$ and $\grs$ are given by \citep{kobayashi97}
\begin{eqnarray}
\gfs &\simeq& \gm \sqrt{ (1 + \frac{2\gm}{\Ga_1} ) / ( 2+\frac{\gm}{\Ga_1})} \\
\grs &\simeq& \gm \sqrt{ (1 + \frac{2\gm}{\Ga_2} ) / ( 2+\frac{\gm}{\Ga_2}) } .
\end{eqnarray}
We note that Equations (6) and (7) are for the case of relativistic shocks, 
where the adiabatic index is 4/3.
For non-relativistic shocks, the adiabatic index should be 5/3,
but since this does not make a large difference, we use this formula
for all shocks in this paper.
Following \citet{kobayashi97}, we estimate the shock crossing time 
$\tcrs$ by the longer time scale of the two shocks to cross the 
shell; this happens to be the time which the reverse shock takes 
to cross the faster shell.  Because the emitting region is moving 
with a relativistic speed towards the observer, the observed duration 
of the flare will be shortened by a factor of 
$(1-\beta \cos \theta ) \sim 1/\gm^2$, and thus,
\begin{equation}
t_{\rm crs} = \frac{l}{c~(\beta_2 - \betars)~\gm^2 } 
	\sim \frac{2}{c~\gm^2} \left( \frac{1}{\grs^2} - \frac{1}{\Ga_2^2} \right)^{-1} ~D_0
\label{eq:d_is:tcrs}   
\end{equation}
where $\beta_2 c$ and $\betars c$ are respectively the velocity of the 
catching up shell and the reverse shock. 
The angular spreading time is given by 
\begin{equation}
t_{\rm ang} = \frac{R}{c~\gm}.
\label{eq:d_is:tang}   
\end{equation}
The average time cycle of the collisions is determined 
from the frequency of the ejection of the shells, and thus
\begin{equation}
\tcol = \frac{4}{c} ~D_0.
\label{eq:d_is:tcol}
\end{equation}
For the $\tshot$ in the simulations, we use the longer one of the 
$\tcrs$ and $\tang$ for each collision.  As a result,
since $\tcrs$ is always 
longer than $\tang$ for all cases, thus $\tshot$=$\tcrs$.

\subsection{Results and the Dependence on Each Parameter}
\begin{figure}
\epsscale{0.5}
\plotone{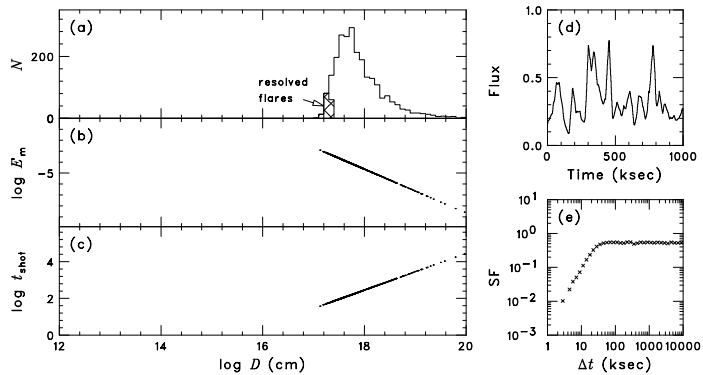}
\figcaption[]{
The results of a simulation of the internal shock model for the case
of $\gav$=10, $\sgd$=0.005, and $D_0$=$3\times10^{13}$ cm.
(a) The histogram of the collision distances $D$.
(b) The dissipated energy 
and the (c) the generated time scale of the flare
at the collision, showed as a function of $D$.
(d) The resulting simulated light curve, and (e) the
structure function calculated from the light curve.
It is apparent that the simulated 
light curve shows repeating resolved flares, 
and that the structure function clearly shows a 
characteristic time scale.
\label{fig:d_is:ini}
}
\end{figure}

We first consider the case of $\gav$=10, $\sgd$=0.05,
and $D_0$=$3\times10^{13}$ cm.  The distribution of the 
collision distances in this case is shown in Figure~\ref{fig:d_is:ini}(a).
Our simulations show that the collisions take place at 
distances ranging from $D\sim10^{17}$ cm up to $D\sim10^{20}$ cm.
Figure~\ref{fig:d_is:ini}(b) shows the amount of the 
newly generated internal energy for each collision $\Em$
plotted against the collision distance $D$, and
Figure~\ref{fig:d_is:ini}(c) is the time scale
of each generated flare $\tshot$ plotted against $D$.
It is apparent that $\Em$ is larger and $\tshot$ is shorter
for collisions which occur at smaller $D$.

A portion of the simulated light curve for this case
is shown in Figure~\ref{fig:d_is:ini}(d).  
It can be seen that the overall light curve is characterized by
repeating flares having time scales of  $\sim 50 - 100$ ks,
resembling the light curves observed from TeV blazars.
This results from the fact that the shells which collide at
smaller distances have larger $\Em$ and shorter $\tshot$; 
with this, the amplitude of the emission becomes much larger 
compared to the shots generated from collisions at larger distances.
Accordingly, only the shots produced by collisions at the smallest
distances will be apparent as flares in the observed light curve.
The average frequency of the collision in this case 
is $F_{\rm col} = 0.24$ mHz, corresponding to one collision 
per $\sim$4 ks on average, while the number of flares which 
are observed is only about 6\% of the total number of shots.
The distribution of the resolved flares are 
shown as the shadowed area in Figure~\ref{fig:d_is:ini}(a).

The calculated structure function for the simulated light curve 
is shown in Figure~\ref{fig:d_is:ini}(e).  A clear break, indicating 
a characteristic time scale $\tchr$ is seen, and the slope at the 
shorter time scales is $\beta \sim$2.  This indicates that the values 
of $\tshot$ of observable flares are restricted to a rather narrow 
range, which determines $\tchr$, and that there is very little 
variability power below $\tchr$ --  exactly what we have observed 
from TeV blazars.

Another remark regards the offset-like component.  
This is due to the other 94\% of the collisions
that generate the longer, smaller amplitude shots,
which overlap each other, resulting in the observed offset. 
In the following, we show how each of the parameters 
affects on these results.

\subsubsection{Dependence on $\gav$}
\label{sec:gav}
We first consider the effect of the change in the average 
BLF $\gav$, while the relative width $\sgd$,
and $D_0$ are kept constant.  When $\gav$ increases by a factor of $a$, 
Equation~\ref{eq:d_is:d} implies that the collisions will occur at 
a distance $D$ which is greater, by a factor of $a^2$.
On the other hand, Equation~\ref{eq:d_is:tcrs} shows that the 
the duration of each shot will be conserved, and thus only the 
amplitude will be different in the resulting light curve.

\subsubsection{Dependence on $D_0$ }
\label{sec:d0}
We then consider the case where the initial separation of the 
colliding shells $D_0$ changes, while $\gav$ and $\sg$ are kept 
constant.  When $D_0$ is larger by a factor of $a$, 
Equations~\ref{eq:d_is:tcrs} and \ref{eq:d_is:tcol}
imply that both $\tcrs$(=$\tshot$) and $\tcol$ will be longer
by a factor of $a$, and thus the simulated light curve will simply 
be a stretched-in-time version of the original one 
-- the amount of relative offset components will be identical.
The collision distance $D$ will become larger by a factor of $a$.

\subsubsection{Dependence on $\sgd$}
\label{sec:sgd}
This turns out to be the most important parameter.
Here, we fix $\gav$=10 and $D_0$=$3\times10^{13}$ cm,
and simulate light curves with different $\sgd$.
First, it can be shown from Equation~\ref{eq:d_is:d} that
as $\sgd$ becomes larger, the collisions start to take place 
at shorter distances from the core.  

\begin{figure}
\epsscale{0.4}
\plotone{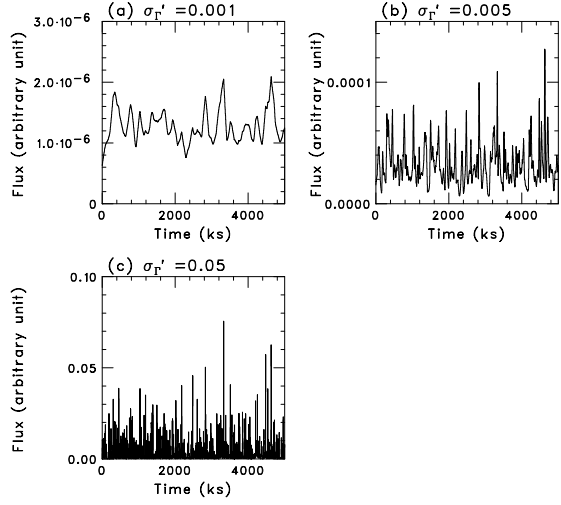}
\figcaption[]{
The simulated light curves for the case
of $\sgd$= (a) 0.001, (b) 0.005, and (c) 0.05.
$\gav$ is fixed to 10, and $D_0$ is fixed to $3\times10^{13}$ cm.
It is shown that the relative amplitude of the flare to
offset component increases as $\sgd$ becomes larger.
\label{fig:d_is:sim3lc}
}
\end{figure}

\begin{figure}
\epsscale{0.4}
\plotone{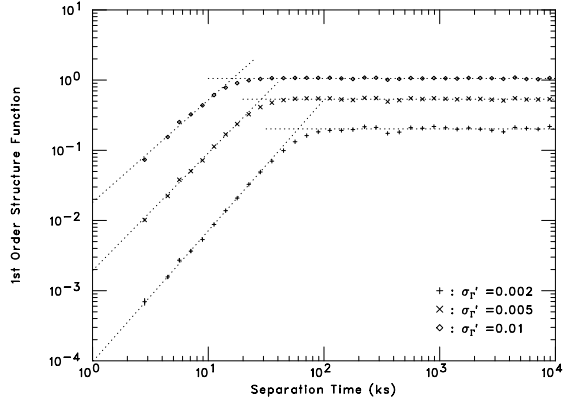}
\figcaption[]{
The structure function calculated from  the
simulated light curves for the case of
$\sgd$=0.002, 0.005, and 0.01.
$\gav$ is fixed to 10, and $D_0$ is fixed to $3\times10^{13}$ cm.
The location of the break, indicating the characteristic time scale, 
shifts to the shorter time scale as $\sgd$ increases.
\label{fig:d_is:sim3sf}}
\end{figure}

Since the characteristic time scale $\tchr$ is always determined by 
the shots due to collisions which took place at the
smallest distances, the time scales of the observable flares in
the light curve become shorter when $\sgd$ is larger.
This is shown in the simulated light curve for different 
$\sgd$ ($\sgd$=0.001, 0.005, and 0.05) in Figure~\ref{fig:d_is:sim3lc}.
Here, the flares appear more spiky as $\sgd$ becomes larger.
Note that since all 3 simulations assume the same $D_0$, 
the number of collisions per unit time is the same for 
all 3 light curves. On the other hand, the number of 
visible flares in the light curves is clearly different.
In order to see the difference in $\tchr$,
we calculated the structure function for the simulated 
light curves. This is shown in Figure \ref{fig:d_is:sim3sf}.
The break, indicating $\tchr$, is clearly seen to
shift to the shorter time scales as $\sgd$ becomes larger.

The differences in the spikiness in the light curve can be regarded 
as the differences in the relative amplitude of the flare and 
offset components.
Importantly, what we have shown is that this relative amplitude, $\rfo$,
changes with the value of $\sgd$. This indicates that,
assuming that all of the offset component is generated 
by the internal shocks, we can compare the observed $\rfo$
to the that derived through simulations. 
Indeed, the actual observations have clearly shown a larger 
offset component than, for instance is apparent 
in Figure~\ref{fig:d_is:sim3lc}(c), which suggests that $\sgd$ must be 
relatively small. We will quantify this in section 4.

\subsubsection{Effect of Fluctuation in $l$}
We have so far fixed the shell thickness $l$ to be equal to the 
initial separation of the two shells $D_0$.  Here at the end of 
this section, we consider the effect when $l$ also fluctuates.
We assume that $l$ is distributed around an average $D_0$,
following a Gaussian function with its width described by the 
relative sigma, $\sld$ ($\equiv \sigma_l / D_0$).  We also make an 
assumption that the mass of the ejected shells scales with the 
shell thickness (i.e. density of shell is constant).

\begin{figure}
\epsscale{0.4}
\plotone{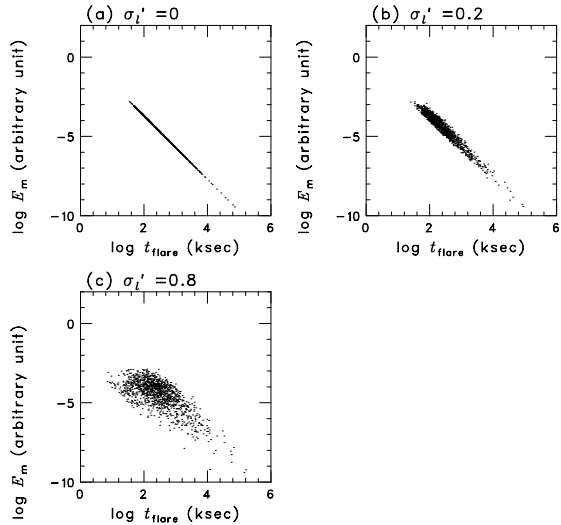}
\figcaption{The correlation of the flare time scale and 
the energy dissipated in each collision for the
case of $\sld$= (a) 0.0, (b) 0.2, (c) 0.8.
The other parameters are fixed to
$\gav$=10, $\sgd$=0.005, and $D_0$=$3\times10^{13}$ cm.
\label{fig:d_is:simlte}}
\end{figure}

The effect of the fluctuation is demonstrated in the plotted 
correlation of the duration of the radiation ($\tshot$) and 
the dissipated energy ($\Em$) for each collision in 
Figure~\ref{fig:d_is:simlte}.  For the case where $l$ is fixed (a), 
$\tshot$ and $\Em$ become simply respectively shorter and larger 
as $\dg$ ($\equiv \Gamma_2 - \Gamma_1$) of the colliding shells increases,
indicating the clear trend in the correlation.  When $l$ fluctuates, 
the spread becomes wider, as shown in Figures~\ref{fig:d_is:simlte}(b)(c).

\begin{figure}
\epsscale{0.4}
\plotone{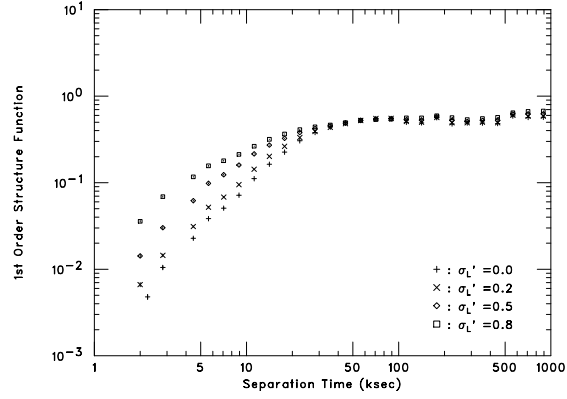}
\figcaption{The structure function calculated from the
simulated light curves for the case of
$\sld$=0.0, 0.2, 0.5, 0.8.
The other parameters are fixed to
$\gav$=10, $\sgd$=0.005, and $D_0$=$3\times10^{13}$ cm.
The slope of the derived structure function flattens as $\sld$ increases.
\label{fig:d_is:simlsf}}
\end{figure}

This indicates that there will be increasing power in the faster 
variability time scales when $\sld$ becomes larger.  For a shell with 
smaller $l$, $\Em$ will be reduced because of the smaller mass of the 
shell.  However, since the shock crossing time $\tcrs$ (=$\tshot$) 
also becomes  shorter, the {\sl amplitude} of the shot will be as large 
as that of a longer shot due to a larger $l$.  This is demonstrated 
in the calculated structure function from the light curves 
simulated for different $\sld$, shown in Figure~\ref{fig:d_is:simlsf}.
While the break is seen to stay nearly at the same value, the slope at 
the faster time scale appears to flatten as $\sld$ increases.

\subsection{Summary of the Simulations} 
The simulation results can be summarized as follows:
\begin{enumerate}
 \item  Collisions of two shells which 	
	had the largest relative velocity,
	and accordingly collided
	at the shortest distances ($D$),
	are the shots which appear as the strongest observable flares
 	in the light curve.
 \item  These shots 
	determine the characteristic time scale ($\tchr$) of the 
	variability. There is very little 
	variability power on time scales shorter than this $\tchr$.
 \item	An offset component will arise from 
	emission due to overlapping shots produced by 
	collisions at larger distances.
	The relative amplitude of flare to offset ($\rfo$)
	is a function of the initial width of the 
	bulk Lorentz factor ($\sgd$).
 \item The dependences of each parameter are:
\begin{itemize}	
\item $\gav$ determines $\Em$:  
      higher $\Em$ yields higher $\gav$.
\item $D_0$ determines the normalization of the time series:  
     longer $\tchr$ is a result of larger $D_0$.   
\item $\sgd$ has an effect on $\tchr$ and $\rfo$:  
    larger $\sgd$ results in smaller $\tchr$ and larger $\rfo$.
\item $\sld$ has effect on the slope $\beta$ of the 
	structure function:  
	$\beta$ flattens as $\sld$ increases.
\end{itemize}

\end{enumerate}

\section{APPLICATION OF THE INTERNAL SHOCK MODEL TO THE X-RAY DATA FOR TEV BLAZARS}
\subsection{Light Curves}
\label{sec:interpretation_lc}
In the previous section, we have shown that the internal shock 
scenario naturally predicts the main features of blazar light 
curves described in \S\ref{sec:obs}, under the condition of $\sgd \ll 0.01$.
One prediction is that the typical time scale of the observed flares 
($\tchr$) always becomes similar to the time scale of the flare $cycles$.
This is very much consistent with the actual observations, where 
the ``day-scale'' flares are observed ``daily.''

The next prediction concerns the features observed in the structure 
function analysis. The observations show that all TeV blazars show 
a break in the structure function, with $\tchr \sim 1$ day. 
The slope below this $\tchr$ is steep ($>$1), suggesting that 
very little variability power is below this $\tchr$.
We have shown that the structure functions calculated from
the simulated light curves show the same features.
In fact, this provides an explanation for the non-existence 
of shorter time scale variability -- there are no 
collisions until a certain distance $D$ from the central core.

\begin{figure}
\epsscale{0.4}
\plotone{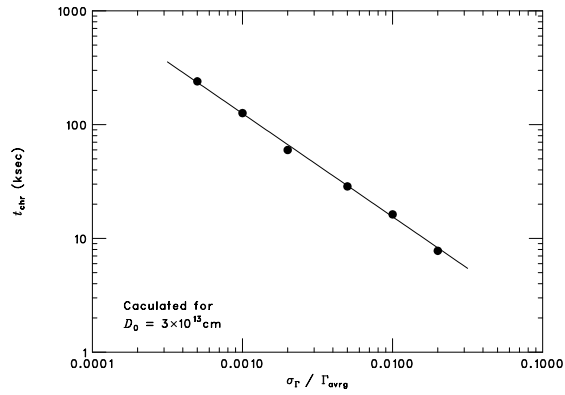}
\figcaption{The derived characteristic times scales of the 
variability plotted as a function of the initial width of the
BLF distribution, $\sgd$.
The case of $D_0$=\ten{3}{13} is shown.
\label{fig:d_is:simtchr}}
\end{figure}

In order to see the actual $\sgd$ dependence on $\tchr$,
we simulated light curves for a series of $\sgd$
for the case of  $\gav$=10, and $D_0$=\ten{3}{13} cm
as in \S\ref{sec:sgd}.
We calculated the structure function for each simulated light curve, 
and assumed that it is well described by two power laws.  We then 
fitted for the two slopes, and estimated the characteristic
time scale as the point where the two slopes cross 
(as shown as the dotted lines in Figure~\ref{fig:d_is:sim3sf}).
The derived values of $\tchr$ plotted as a function of $\sgd$
are shown in Figure~\ref{fig:d_is:simtchr}.  It appears that 
$\tchr$ becomes shorter with larger $\sgd$,
and the best-fit power-law gives a slope of  $\sim -0.9$.
Note that we have shown in \S\ref{sec:gav} that the time scales do 
not depend on  $\gav$ if $\sgd$ is constant.

Another point regards the presence of the offset component.
As discussed in \S\ref{sec:obs}, the observed flares in TeV 
blazars always appear to lie on top of an underlying offset component. 
The data do not tell us directly whether this offset is due to flares 
overlapping each other, or due to some steady emission component.
What we have shown here is that considering the internal shock scenario,
there will always be many overlapping flares 
which will appear as the offset component.

The amount of this offset component should also be useful in 
modeling of an actual observation, as this suggests that $\sgd$ can 
be estimated from an observed light curve if it has a sufficient 
length to estimate the relative amplitude of the flare and offset 
component.  In an attempt to quantify the offset component, here 
we use the parameter $\rfo$, describing the relative 
amplitude of the flare and offset component as follows.
We first generate a histogram of the count rates which form a peak. 
Since the lower and higher end of the peak represent the minimum 
and maximum count rates in the light curve, these two can be 
considered as an indicator of the amplitude of the offset component, 
and the offset-plus-flare component.  As there are fluctuations, 
we define the offset amplitude as the point where 10\% of total
counts is reached, and the offset-plus-flare amplitude
as the point where 90\% of the total counts is included.
The amplitude of the flare component is estimated by
subtracting the offset amplitude from the offset-plus-flare amplitude.

We calculate this flare-to-offset ratio $\rfo$ for the same set 
of simulated light curves used in Figure~\ref{fig:d_is:simtchr}.
The result is shown in Figure~\ref{fig:d_is:simrfo},
which shows that $\rfo$ increases with increasing $\sgd$.
As we showed in \S\ref{sec:sim}, $\rfo$ does not depend on 
$\gav$ nor $D_0$, which means that $\sgd$ can be directly estimated
from a  $\rfo$ given by an observed light curve.

Finally, we note that the $\tchr$ plotted in the 
Figure~\ref{fig:d_is:simtchr} is for the case of $D_0$=$3\times10^{13}$ cm. 
Given that the time series scales linearly with $D_0$,
the vertical axis in Figure~\ref{fig:d_is:simtchr} can be regarded as 
$\tchr /D_{0,3\times10^{13}}$, where $D_{0,3\times10^{13}}$ 
denotes $D_0$ in units of $3\times10^{13}$ cm.  
Accordingly, this indicates that $D_0$ can also be estimated if
$\rfo$ and $\tchr$ can be measured.
This is interesting, given that the initial separation 
is a value determined from the frequency of the emitted shells.
Thus, $D_0$ should reflect the frequency of the 
activity of the central engine, that generates and ejects 
individual shells. 

\subsection{Variation of the Synchrotron Peak Frequency}
The important observational result from the spectral analysis was 
that the peak synchrotron frequency increases during 
both high intensity states, and also within the daily flares.
In this section, we discuss whether this can be interpreted 
within the same model.  We start with formulating the peak energy, 
by generally following the prescriptions of \citet{inoue96} and \citet{kirk98}.

The peak frequency of the synchrotron spectrum reflects the maximum 
energy of the accelerated electrons, which is determined from where 
the cooling and the acceleration time scale becomes comparable. 
The cooling time of the electrons due to synchrotron
and inverse Compton emission can be written as \citep{rybicki},
\begin{equation}
\tau _{\rm cool}(\gamma) = \frac{3\me c}{4(\Ub + \usoft)\st \gamma} ,
\label{eq:d_is:tcool1}
\end{equation}
where $\Ub$ and $\usoft$ are the energy densities of the
magnetic field and the soft photons to be upscattered 
in the Inverse Compton process, $\me$ the rest mass and and $\gamma$ is the 
random Lorentz factor of the electron, and $\st$ is the Thomson 
cross section.

The acceleration time is not as well understood as the cooling time.
Perhaps the most promising theory suggests that it is determined 
from the time scale of the 1st order Fermi acceleration process 
operating in a shock.  In this case, the acceleration time can be 
approximated by considering the mean free path $\lambda(\gamma)$ for the 
scattering of the electrons with the magnetic disturbances.  Taking 
the mean free path to be proportional to the Larmor radius by 
introducing another parameter $\xi$ 
($\lambda \equiv \xi \frac{\gamma mc^2}{eB}$),
the acceleration time is given by,
\begin{equation}
\tau _{\rm acc}(\gamma) = \frac{20\lambda(\gamma)c}{3\vs^2} 
	= \frac{20 \me c^3 }{3eB} \frac{\xi}{\vs^2} \gamma,
\label{eq:d_is:tacc}
\end{equation}
where $\vs$ is the shock velocity.
By equating the radiative cooling and acceleration times 
in Equations \ref{eq:d_is:tcool1} and \ref{eq:d_is:tacc}, the
maximum energy of the electrons is,
\begin{equation}
\gmax = \frac{\vs}{c} 
	\left( \frac{9eB}{80(\Ub+\usoft)\st\xi} \right) ^{\frac{1}{2}}.
\label{eq:d_is:gmax}
\end{equation}
Furthermore, in the case of TeV blazars,
the Compton cooling for the energy range of electron
emitting X--rays is strongly suppressed by the Klein-Nishina cutoff, 
and thus $\usoft /\Ub \ll 1$ (e.g. Kataoka 2000; Li \& Kusunose 2000).
Thus for this case, the observed maximum synchrotron frequency 
can be written as 
\begin{eqnarray}
\nmax &=& 1.2\times 10^6 B \delta \gmax^2 \\
      &\simeq& 2.49 \times 10^{21} \frac{\delta_{10}}{\xi} 
	\left( \frac{\vs}{c} \right) ^2 {\rm Hz},
\label{eq:d_is:nmax} 
\end{eqnarray}
where $\delta_{10}$ denotes the beaming factor in units of 10.
Note that the above implies that when synchrotron losses dominate
(as is likely to be the case for TeV blazars), 
the observed peak frequency is independent on $B$.

We will first consider the differences in the synchrotron spectrum
for the flare components and the underlying offset component.
Above, we have shown that in the framework of the internal shock 
scenario, the observed flares can be regarded as being due to collisions 
which had the largest $\dg$ within the distribution, while 
the offset component results from emission due to fainter, overlapping 
flares produced by collisions that had smaller $\dg$.

Since the collisions which generate the observed flares 
occur at shorter distances $D$, the magnetic field $B$ is 
expected to be higher.  Accordingly, the radiative cooling 
times of the electrons are shorter there, preventing electrons from 
being accelerated to higher energies.  On the other hand, 
Equation~\ref{eq:d_is:tacc} suggests that the acceleration time 
will also be shorter, due to the larger $B$, and also due to larger 
shock velocity $\vs$.  Equation~\ref{eq:d_is:nmax} shows that 
concerning the observed synchrotron peak {\it frequency},
the changes in $B$ cancel out and only values of $\vs$, $\delta$, 
and $\xi$ affect $\nmax$. 

\begin{figure}
\epsscale{0.35}
\plotone{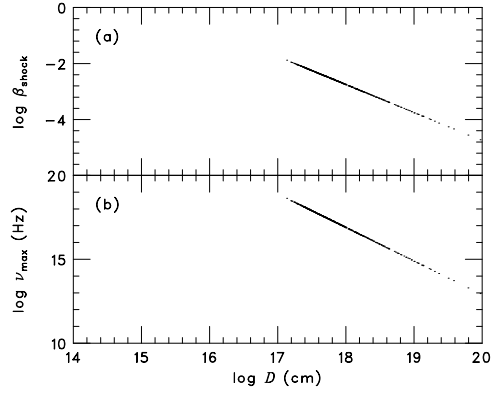}
\figcaption{The relative velocity of the reverse shock (a)
and the resulting synchrotron peak frequency (b) calculated for 
the case of $\gav$=10, $\sgd$=0.005, and $D_0$=$3\times10^{13}$ cm.
The Bohm limit is assumed.  
\label{fig:d_is:ini_vs}}
\end{figure}

In Figure~\ref{fig:d_is:ini_vs}(a), we show the 
calculated shock velocity $\vs$ for each collision 
(in units of $c$; $\bshock$=$\vs /c$) as a function of $D$,
for the case of $\gav$=10, $\sgd$=0.05,
and $D_0$=$3\times10^{13}$ cm (similar to the
ones used in Figure~\ref{fig:d_is:ini}). 
It shows that $\vs$ decreases as $D$ increases, following a relation of
$\bshock \propto D^{-1}$.  Accordingly, if $\xi$ were to be similar for 
all collisions, Equation~\ref{eq:d_is:nmax} shows that $\nmax$ will be 
higher for collisions at smaller $D$.  Since these are the shots that 
appear as flares, this is consistent with the actual observations,
where the flare spectrum components show higher $\nmax$ than the 
offset component.  The calculated $\nmax$, considering the Bohm limit 
(i.e. $\xi$=1) is shown in Figure~\ref{fig:d_is:ini_vs}(b).
The dependence of $\nmax$ on $D$ is $\nmax \propto D^{-2}$.

Finally, in addition to the spectral variation during the day-scale flares, 
the same trend regarding  the relationship between the 
flux and synchrotron peak frequency is also observed in comparing 
different observations.
The same discussions as above will hold for the case when
$\sgd$ becomes larger. 
The shock velocity will become higher for the collisions
which generate the observed flares, and accordingly the average
$\nmax$ will be higher.
Another way to generate a change in the average $\nmax$
is the change in the $\gav$.
If $\gav$ increases, it can be shown from equation~\ref{eq:d_is:nmax}
that $\nmax$ will also increase.

\subsection{The Case of Mrk~421}
\label{sec:d_is:mrk421-1}
Here we consider the observed light curve of Mrk~421
during the \asca\ long look in 1998 April.  
In modeling the observed light curve, we wish to reproduce 
the observables determined from these data such as 
the characteristic time scale $\tchr$ and the 
flare-to-offset ratio $\rfo$ (as defined 
in \S~\ref{sec:interpretation_lc}).

\begin{figure}
\epsscale{0.4}
\plotone{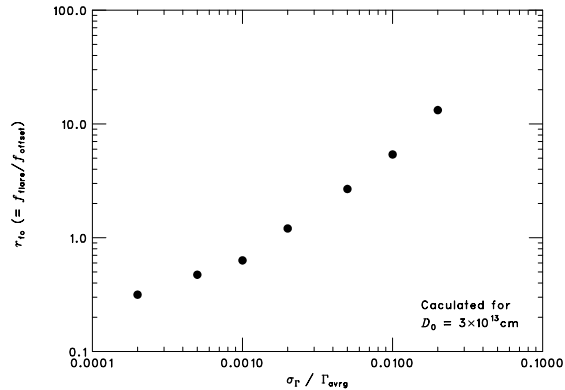}
\figcaption{The ratio of the flare to offset component amplitude, 
plotted as a function of the initial width of the 
BLF distribution, $\sgd$.
\label{fig:d_is:simrfo}}
\end{figure}

\begin{figure}
\epsscale{0.4}
\plotone{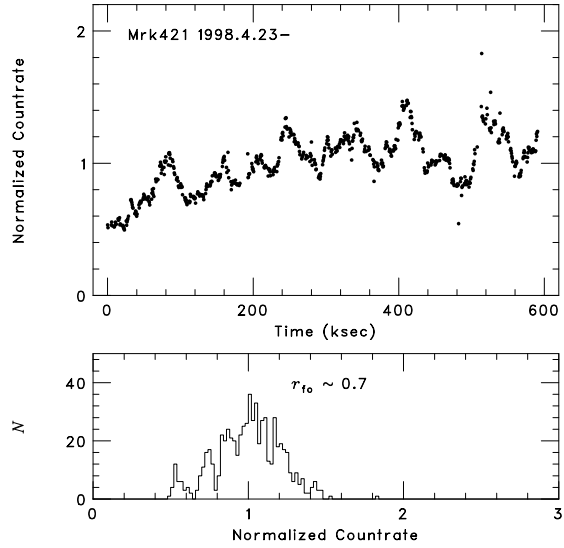}
\figcaption{The observed normalized 0.6--7.5 keV count rate light 
curve and the histogram of the count rate 
for Mrk~421 during the \asca\ long look in 1998 April.
The bottom panel shows the number histogram of the count rate,
where the lower and higher end of the peak can be considered as 
an indicator of the amplitude of the offset component, and 
the offset-plus-flare component.
The derived flare-to-offset ratio defined in \S4.1 is
calculated to be $\rfo \sim$0.7 for this case.
\label{fig:d_is:421hist}}
\end{figure}

The first step is to estimate $\sgd$ from the observed $\rfo$.
The normalized light curve (in counts s$^{-1}$) and the calculated 
histograms of the count rate are shown in Figure~\ref{fig:d_is:421hist}.
The calculated $\rfo$=0.7 is compared with the derived relation of
$\sgd$ and $\rfo$  from the simulations shown in 
Figure~\ref{fig:d_is:simrfo}, which gives an estimate of 
$\sgd \sim$0.001.  Then, from the observed characteristic time scale
$\tchr \sim 40$ ks \citep{tanihata01}, $D_0$ can be estimated
from Figure~\ref{fig:d_is:simtchr} as
\begin{equation}
D_0 \sim \frac{40}{120} \times 3\times 10^{13} \sim 1\times 10^{13} {\rm cm}.
\end{equation}
From Equation~\ref{eq:d_is:tcol}, this would give $\tcol \sim 1300$ seconds,
which indicates that the shells are ejected from the central engine 
on average every $\sim$10 minutes.  

\begin{figure}
\epsscale{0.4}
\plotone{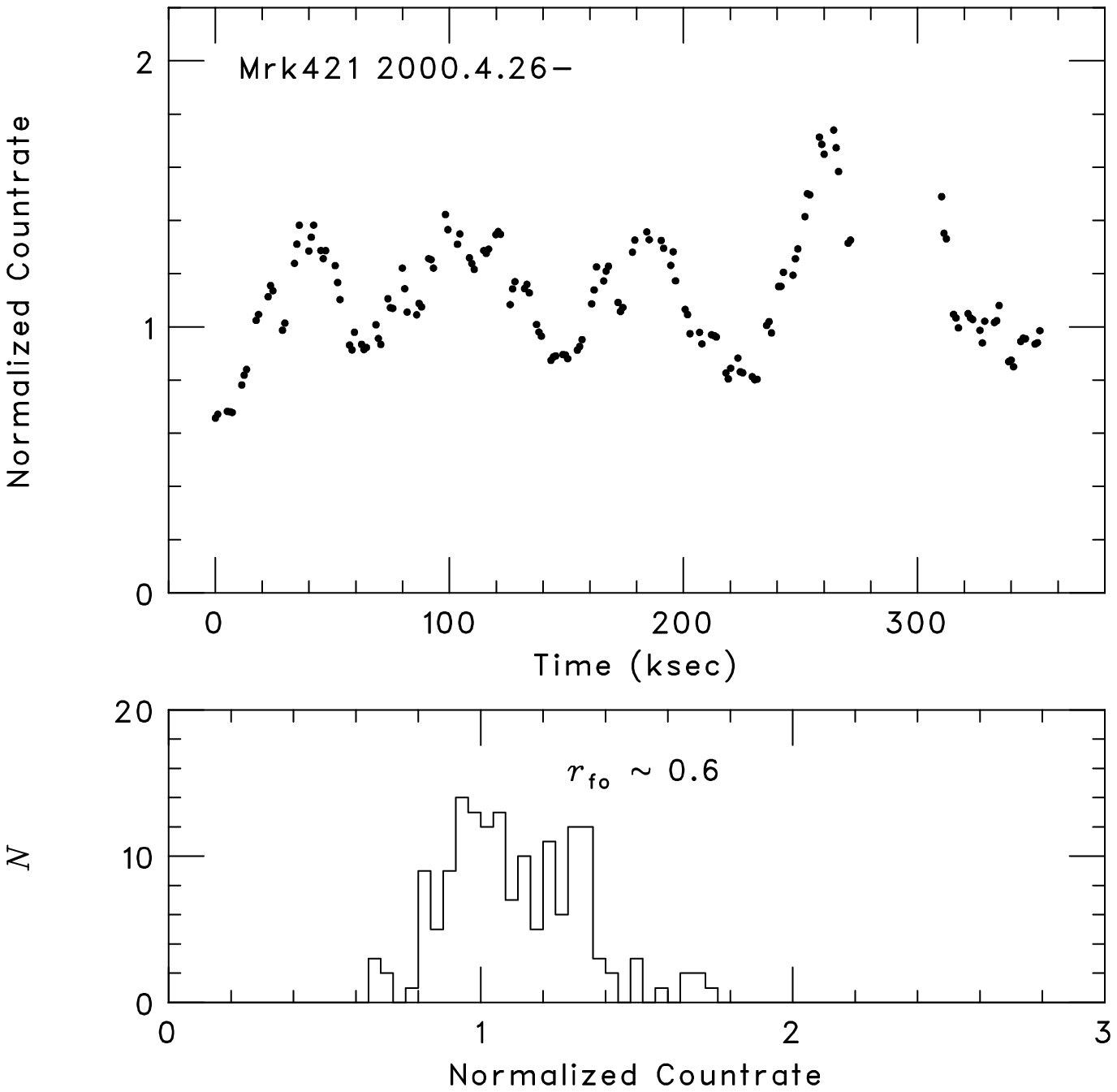}
\figcaption{The observed normalized count rate light curve and the
histogram of the count rate for the \sax\ observation of Mrk~421 
during 2000 April.  The light curve was generated by integrating 
over the MECS band.  The bottom panel shows the number histogram 
of the count rate, where $\rfo \sim$0.6 is derived.
\label{fig:d_is:421histsax1}}
\end{figure}

\begin{figure}
\epsscale{0.4}
\plotone{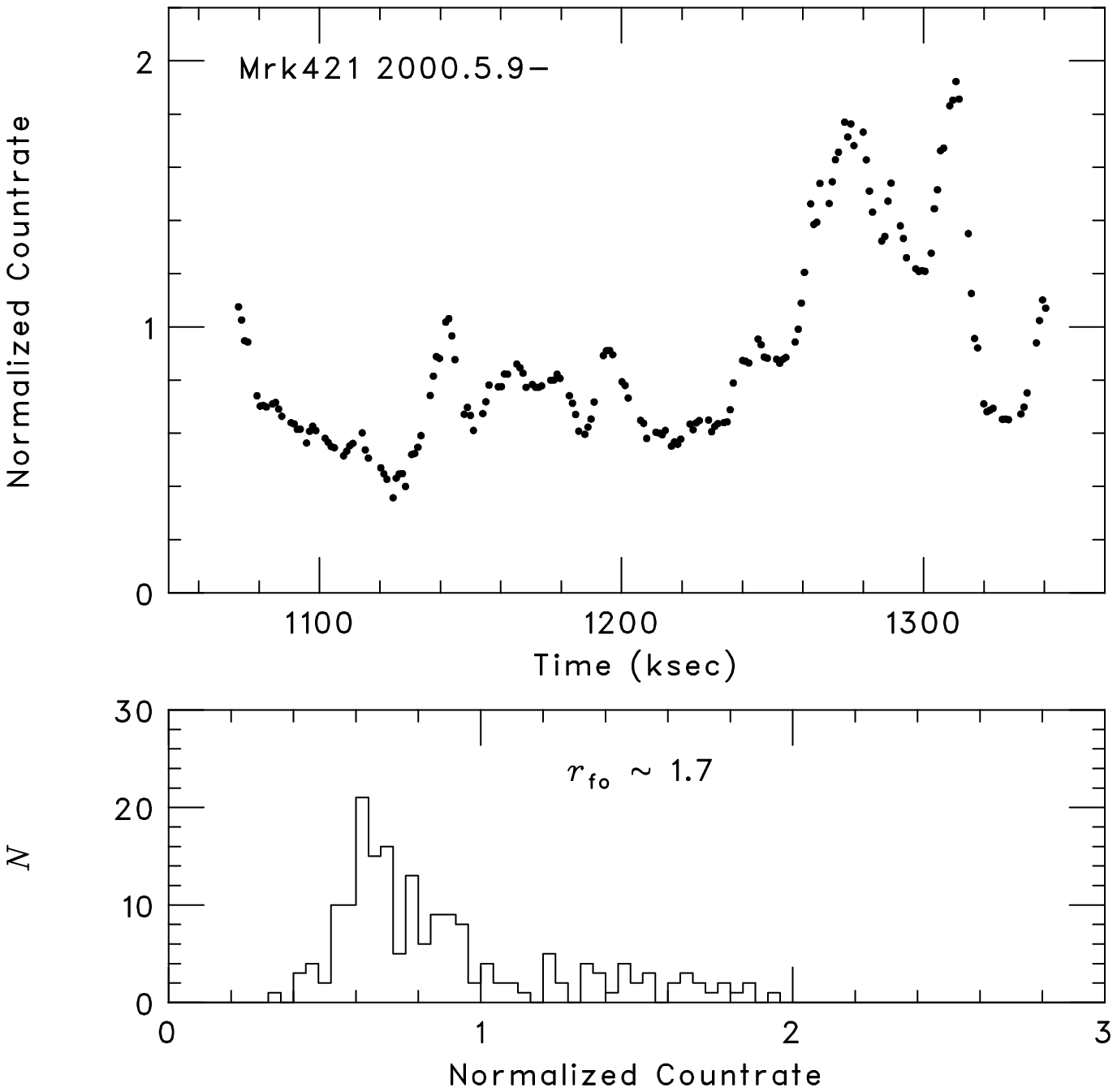}
\figcaption{The observed normalized count rate light curve and the
histogram of the count rate 
for the \sax\ observation of Mrk~421 during 2000 May.
The light curve was generated by integrating over the MECS band.
The bottom panel shows the number histogram of the count rate,
where $\rfo \sim$1.7 is derived.
\label{fig:d_is:421histsax2}}
\end{figure}

\begin{figure}
\epsscale{0.4}
\plotone{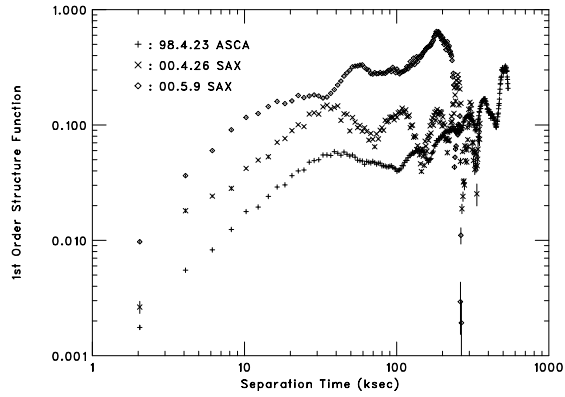}
\figcaption[]{The structure function calculated for the
observed light curve of Mrk~421 for the 3 observations shown in 
Figures~\ref{fig:d_is:421hist}-\ref{fig:d_is:421histsax2}.
\label{fig:d_is:421sf}
}
\end{figure}

On the other hand, we also remark here that the $\rfo$ derived 
in Figure~\ref{fig:d_is:simrfo} is for the variations of the 
dissipated energy.  Concerning the $\rfo$ for a particular energy,
this will have an energy dependence, which requires
a proper treatment of synchrotron formulae, and will also
depend on the energy dependence of the efficiency 
of the energy that is to be converted  to radiation. 
This might be further investigated by studying the difference of  
$\rfo$ between different observations. In particular, a similar 
analysis as above, applied to the \sax\ observation of Mrk~421 
in 2000 May, shows a larger value of $\rfo$=1.7,
whereas the value is somewhat similar to the
\asca\ observations for the \sax\ observations in 2000 April.
This is shown in Figure~\ref{fig:d_is:421histsax1} and 
\ref{fig:d_is:421histsax2}.  An examination of the light curve 
during 2000 May reveals that flares with shorter time scales are 
dominant in the light curve.  These differences in the characteristic 
time scales can be seen clearly in the structure functions calculated 
for each observation, shown in Figure~\ref{fig:d_is:421sf}.  The break 
in the structure function is at $\sim$10 ks for the 2000 May data,
whereas it is at $\sim$30--40 ks for the other two.  While this might 
be only a coincidence, the 3 observations show a trend that 
$\rfo$ is larger when $\tchr$ is shorter:  this is actually the trend 
expected from our model -- when $\sgd$ is larger, $\rfo$ is larger 
and $\tchr$ is shorter.

\subsection{The Efficiency}
One concern is that the dynamical efficiency $\eta$ of the collisions 
is rather small.  For instance, for the parameters derived above
for Mrk~421, $\eta$ given by Equation~\ref{eq:d_is:eta}
is $\sim 10^{-6}$ for the collisions that occur at the shortest $D$.
For the more distant collisions, this becomes even smaller.

It has been claimed that the jet cannot radiate all of its energy 
in the sub-parsec region considered here, since a substantial power 
must be transported to the kpc-scale radio lobes.  With this, the 
efficiency must be low.  However, this seem too low -- as this means that
only 10$^{-6}$ of the total energy 
is radiated at the base of the jet, which 
indicates that the jet is 10$^6$ times energetic than is observed 
via the blazar phenomenon.  On the other hand, since there is no measure 
of the energetics of the entire jet, such low radiative efficiency near 
the core  cannot be  simply ruled out.  Nevertheless, as it does seem 
rather low, we discuss how this could be increased.

The very low efficiency mainly results from the fact that we intend to 
reproduce the offset component, which reduces the $\sgd$,
and thus reduces the velocity difference of the two colliding shells. 
Thus, the easiest way to increase the efficiency would
be to have larger $\Ga_2 -\Ga_1$, but this would make the
flare time scales shorter than the flare cycles, which conflicts 
with the observed flares that occur repeatedly (daily-flares).

Part of the low efficiency also is due to the fact that 
we are assuming a Gaussian distribution for the initial BLF,
which emphasizes the effect.  
The origin of the modulation of the Lorentz factor is not 
known; it can be due to any physical condition 
of the central engine that is not stable,
such as instabilities in the innermost parts of an accretion disk or
magnetic eruptions in the corona (e.g., Sikora \& Madejski 2000).
Thus one way to increase the efficiency 
would be to assume a broader distribution, such as a flat distribution 
within a certain range (e.g., Spada et al.\ 2001; Kobayashi, Piran, \& 
Sari 1997), 
which would increase the number of efficient collisions.
According to our simulations, this was shown to increase the efficiency 
by roughly one order of magnitude.  We can consider even more extreme 
distributions where the values of the BLFs are concentrated at the low 
and high end of the distribution.  At most, this could improve the 
efficiency by another order of magnitude, but not any more, since 
$\Ga_2 -\Ga_1$ is always small.  If this is still too inefficient,
this would indicate that there is a  problem assuming random distribution 
of BLFs for the shells in the internal shock scenario.  In this case, 
we would probably need to consider another possible origin of the 
offset component.  
For instance, if there is emission from significantly larger distances,
such as by reconfinement shocks or
by shocks due to collisions with inhomogeneities that have
much larger radial extensions,
this may produce a low-amplitude relatively steady component.
For this case, the efficiency can be much larger.
On the other hand, we emphasize here that the internal shock scenario alone
reproduces successfully many of the observed features such as the 
daily-flares, the existence of the characteristic time scale $\tchr$ 
and the non-existence of variability power below $\tchr$.  

\section{THE EXTERNAL SHOCK SCENARIO}
In the previous sections we have discussed how the internal shock model 
successfully reproduces the observed variability of the both the 
fluxes and spectra of blazars.  While the this model works 
rather nicely naturally explaining the observed properties,
we briefly consider the alternative model, the external shock scenario.

In the internal shock scenario, a shock is generated when a faster 
shell catches up to a slower shell. This is where electrons are 
accelerated to relativistic energies and radiate through 
synchrotron and inverse Compton emission.  The alternative scenario 
is that these shells do not collide with each other but run into 
either an external material or external field, where the shock emerges
(e.g. Dermer et al. 1999).  This is somewhat similar to the 
acceleration mechanism which is usually considered in 
supernova remnants or the afterglows of gamma-ray bursts.
In these two examples, the external material is provided by the 
interstellar medium, while in the case of blazars, the promising 
candidates for the external medium are broad line clouds and/or 
intercloud material.  We will discuss below how this may work in 
blazars;  below, we will call such external material ``clumps.''

Here, we assume a shell ejected with a Lorentz factor $\Ga$, 
which runs into a broad line cloud ``clumps'' (or any other external material) 
at a distance $D$.  In similarity to the case of the internal shock 
scenario, given that the radiation cooling time is much faster than 
the observed variability time scales, the observed variability time scales
are most likely determined by the dynamical time scale.  
There are two time scales which may affect the dynamical time scale:  
the time for the shock to cross the region, and the angular spreading 
time.  Here, in contrast to the internal shock scenario, the shock can 
be considered as relativistic.  

For this case, the time for the shock to cross the shell is given by 
\begin{equation}
\tcrs = \frac{l}{2c \Ga^2} ,
\label{eq:d_is:tcrs_ext}
\end{equation}
where $l$ is the shell thickness.  The angular spreading time is 
similar to the case of the internal shock model, and thus
\begin{equation}
\tang = \frac{R}{c \Ga} = \frac{D}{c \Ga^2},
\label{eq:d_is:tang_ext}
\end{equation}
where $R$ is the radius of the shell at distance $D$.

If the shock crossing time $\tcrs$ were to determine the observed time 
scale of the flares, which is of the order of a day,
Equation~\ref{eq:d_is:tcrs_ext} would suggest that the thickness of the 
shell must be as large as $l\sim$\ten{3}{17} cm.  This means that the 
central engine must be continuously ejecting material for 100 days to 
produce a single shell -- which is in severe conflict with the observed 
rate of flares which occur daily.  Thus, in contrast to the case 
of the internal shocks, the angular spreading time $\tang$ should be 
the dominating time scale in the case of external shocks.  For this case, 
to obtain flares with time scales of $\sim$ 1 day, the location of the 
shock is calculated to be at $D \sim$\ten{3}{17}$\Ga_{10}^2$ cm,
where $\Ga_{10}$=$\Ga /10$. 

\begin{figure}
\epsscale{0.4}
\plotone{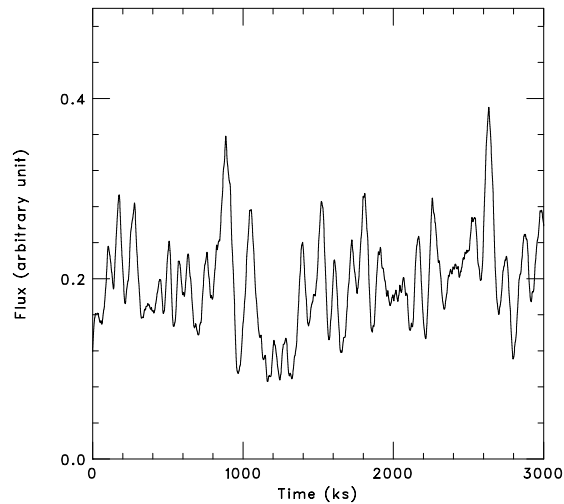}
\figcaption{The simulated light curve considering the external 
shock scenario, assuming parameters $\Ga$=15 and $D$=\ten{5.4}{17}.
\label{fig:d_is:extlc}}
\end{figure}

Because the emission is beamed, the observed time scale of $\sim 1$ day 
reflects a $\sim 10$ day emission in the jet co-moving frame.
The daily occurrence of flares indicates that there must be
at least 10 shells radiating at the same time, which requires 
at least 10 separate ``clumps.''  The external fields must also be 
restricted in a rather limited range, as to keep the flare time scales 
similar.  Figure~\ref{fig:d_is:extlc} shows a simulated light curve for 
the case of $\Ga$=15, and $D$=\ten{5.4}{17} cm
(calculated for $\tshot=\tang=80$ ks).  Here, the average flare cycle 
is set to 5 ks, showing that it is possible for the offset component 
to be produced only if the flare cycle is high (i.e. substantial 
number of ``clumps'' of the external material available for collisions 
with the jet at the same time).

We also remark that for the case of the external shock model, 
the collisions are likely to occur farther away from the central source
than for the internal shock model.  A concern arises to whether 
the broad line clouds are still present at such distances. From 
observations of the time delay between the continuum and line 
variations in AGNs, the broad line regions are suggested to
extend to distances around 10$^{16-18}$ cm, although there are no 
measurements for blazars. If this is of the same order for blazars,
the distance calculated above (for $\Ga$=15, $\tshot$=1 day)
is actually at the right location, but may be close to the limit.
Summarizing, we do not claim that the discussion above rules out the
external shock scenario, but we remark that more fine-tuning of 
the parameters is necessary as compared with the internal shock model.

\section{SUMMARY}
The high-quality light curves and spectra obtained via the 
recent long look observations of TeV blazars provided the first 
opportunity to use the variability as a new tool to study the 
structure of jets in blazars in more detail than was previously possible.  
Our approach is unique as it investigates not only a single 
flare, but considers a series of flares resulting from well-sampled, 
long-duration time series.  We summarized the observed variability 
properties and suggested a physical interpretation to explain
these features based on the framework of the internal shock scenario.

We presented a simplified model applied specifically for TeV blazars, 
and investigated through simulations how each of the parameters 
would affect the observed light curve or spectrum.  In particular, we 
showed that the internal shock scenario naturally accounts for the 
observed variability properties, but it requires a condition 
that the fluctuation of the initial bulk Lorentz factors are small, 
$\sgd \ll 0.01$.  Remarkably, this explains both features observed in the 
flux but also in the spectral variability -- the repeatedly occurring flares, 
the offset component, the structure function, and the shift of the 
synchrotron peak frequency.  We also showed that several useful observational 
quantities can be used to probe the physical parameters of the inner  
jet:  the characteristic time scale, the flare-to-offset ratio,
and the slope of the structure function.  We applied this model to the 
\asca\ X--ray light curves of 
the TeV blazar Mrk~421, which allows us to determine several physical 
parameters of the jet such as the frequency of ejection of shells (on average 
one shell every 10 minutes), the average thickness of the shell 
($\sim10^{13}$ cm), and the location of their collisions (typically 
$\sim10^{17}$ cm away from the central source).  We also briefly commented 
on the external shock scenario, and claimed that this scenario is viable, 
but requires rather detailed fine-tuning to the parameters.

\acknowledgements
We thank Marek Sikora and Fumio Takahara for valuable 
discussions on this manuscript. We would also like to
thank the anonymous referee for constructive comments to
improve this paper.
Support for this work 
was provided by the Fellowship of Japan Society for Promotion 
of Science for Young Scientists, and by NASA via Chandra grant 
no. GO0-1038A from SAO to Stanford University.  








\begin{thebibliography}{}
\bibitem[Bell(1978)]{bell78} Bell, A.~R.\ 1978, \mnras, 182, 147
\bibitem[Blandford \& Eichler(1987)]{blandford87}
	 Blandford, R.~\& Eichler, D.\ 1987, \physrep, 154, 1
\bibitem[Blandford \& Rees(1978)]{blandford78} 
	Blandford, R.~D.~\& Rees, M.~J.\ 1978, \physscr, 17, 265
\bibitem[Dermer, Schlickeiser, \& Mastichiadis(1992)]{dermer92}
        Dermer, C. D., Schlickeiser, R., 
        \& Mastichiadis, A. 1992, \aap, 256, L27
\bibitem[Dermer \& Schlickeiser(1993)]{dermer93}
        Dermer, C. D., \& Schlickeiser, R. 1993, \apj, 416, 458
\bibitem[Drury(1983)]{drury83}
	Drury, L. O'C. 1983, Rep. Prog. Phys., 46, 973
\bibitem[Fossati et al.(2000a)]{fossati00a} 
	Fossati, G.~et al.\ 2000, \apj, 541, 153 
\bibitem[Fossati et al.(2000b)]{fossati00b} Fossati, G.~et al.\ 
2000, \apj, 541, 166 
\bibitem[Ghisellini \& Maraschi(1989)]{ghisellini89}
        Ghisellini, G., \& Maraschi, L. 1989, \apj, 340, 181
\bibitem[Ghisellini(2001)]{gab01}
	Ghisellini, G.\ 2001, ASP Conf.~Ser.~227: 
	Blazar Demographics and Physics, 85
\bibitem[Hoyle et al.(1966)]{hoyle66}
	Hoyle, F., Burbidge, G. R., \& Sargent, W. L. W.
	1966, Nature, 209, 709
\bibitem[Inoue \& Takahara(1996)]{inoue96}
         Inoue, S., \& Takahara, F. 1996, \apj, 463, 555
\bibitem[Jones \& Ellison(1991)]{jones91} 
	Jones, F.~C.~\& Ellison, D.~C.\ 1991, Space Science Reviews, 58, 259
\bibitem[Jones, O'Dell, \& Stein(1974)]{jones74} 
        Jones, T. W., O'Dell, S. L., \& Stein, W. A. 1974, \apj, 188, 353
\bibitem[Kataoka(2000)]{kataokaD} 
        Kataoka, J., Ph.D Thesis, University of Tokyo 
\bibitem[Kataoka et al.(2001)]{kataoka01} 
	Kataoka, J.~et al.\ 2001, \apj, 560, 659
\bibitem[Kirk, Rieger \& Mastichiadis(1998)]{kirk98}
	Kirk, J. G., Rieger, F. M., \& Mastichiadis, A.
	1998, \aap, 333, 452
\bibitem[Kobayashi, Piran, \& Sari(1997)]{kobayashi97} 
	Kobayashi, S., Piran, T., \& Sari, R.\ 1997, \apj, 490, 92
\bibitem[Longair(1994)]{longair94} 
	Longair, M.~S.\ 1994, Cambridge: Cambridge University Press, 
	|c1994, 2nd ed., 
\bibitem[Pian et al.(1998)]{pian98}
        Pian, E., et al.\ 1998, \apj, 492, L17
\bibitem[Rees(1978)]{rees78} Rees, M.~J.\ 1978, \mnras, 184, 61P
\bibitem[Rybicki \& Lightman(1979)]{rybicki} 
        Rybicki, G. B., \& Lightman, A. P. 1979,
        Radiative Processes in Astrophysics (New York: Wiley)
\bibitem[Sari \& Piran(1995)]{saripiran95} 
	Sari, R.~\& Piran, T.\ 1995, \apjl, 455, L143 
\bibitem[Sikora, Begelman, \& Rees(1994)]{sikora94}
        Sikora, M., Begelman, M. C., \&Rees, M. J. 1994, \apj, 421, 153
\bibitem[Sikora \& Madejski(2000)]{sikora00}
	Sikora, M.,~\& Madejski, G.\ 2000, \apj, 534, 109 
\bibitem[Sikora, B{\l}a{\. z}ejowski, Begelman, \& Moderski(2001)]{sikora01} 
	Sikora, M., B{\l}a{\. z}ejowski, 
	M., Begelman, M.~C., \& Moderski, R.\ 2001, \apj, 561, 1154
\bibitem[Spada et al.(2001)]{spada01}
 	Spada, M., Ghisellini, G., Lazzati, D., 
	\& Celotti, A.\ 2001, \mnras, 325, 1559
\bibitem[Takahashi et al.(2000)]{tad00} 
        Takahashi, T., et al.\ 2000, ApJ, 542 L105
\bibitem[Tanihata et al.(2001)]{tanihata01}
	Tanihata, C.~et al.\ 2001, \apj, 563, 569
\bibitem[Tanihata et al.(2002)]{tanihata02}
	Tanihata, C.~et al.\  in preparation
\bibitem[Tavecchio et al.(2001)]{fab01} 
	Tavecchio, F.~et al.\ 2001, \apj, 554, 725
\bibitem[Urry et al.(1997)]{urry97} 
        Urry, C.~M.~et al.\ 1997, \apj, 486, 799
\bibitem[Zhang et al.(2002)]{zhang02}
	Zhang, Y.~H.,~et al.\ 2002, \apj, 572, 762
\end{thebibliography}
\end{document}